\documentclass[traditabstract]{aa} 
\usepackage{graphicx} 
\usepackage[varg]{txfonts}
\usepackage{natbib}
\usepackage{lscape}
\usepackage{rotating}
\usepackage[normalem]{ulem}
\usepackage{subfig}
\usepackage{caption}

\bibpunct{(}{)}{;}{a}{}{,} 

\begin{document}

   \title{Spectroscopic and photometric studies of white dwarfs in the Hyades}

   \author{P.-E. Tremblay\inst{1}
          \and
          E. Schilbach\inst{2}
          \and
          S. R\"oser\inst{2}
          \and
          S. Jordan\inst{2}
          \and
          H.-G. Ludwig\inst{1}
          \and
          B. Goldman\inst{3}
          }

   \institute{Zentrum f\"ur Astronomie der Universit\"at Heidelberg, Landessternwarte, 
            K\"onigstuhl 12, D-69117 Heidelberg, Germany\\
             \email{ptremblay@lsw.uni-heidelberg.de,hludwig@lsw.uni-heidelberg.de}
         \and
         Zentrum f\"ur Astronomie der Universit\"at Heidelberg, Astronomisches
         Rechen-Institut, M\"onchhofstrasse 12$-$14, D-69120 Heidelberg, Germany\\
             \email{elena@ari.uni-heidelberg.de,roeser@ari.uni-heidelberg.de,jordan@ari.uni-heidelberg.de}
         \and
           Max-Planck-Institut f\"ur Astronomy, K\"onigstuhl 17, D-69117
           Heidelberg, Germany\\
             \email{goldman@mpia.de}
             }

   \date{Received ..; accepted ..}
 
  \abstract {The Hyades cluster is known to harbour ten so-called classical
    white dwarf members. Numerous studies through the years have predicted
    that more than twice this amount of degenerate stars should be associated
    with the cluster. Using the PPMXL catalog of proper motions and positions,
    a recent study proposed 17 new white dwarf candidates. We review the
    membership of these candidates by using published spectroscopic and
    photometric observations, as well as by simulating the contamination from
    field white dwarfs. In addition to the ten classical Hyades white dwarfs,
    we find six white dwarfs that may be of Hyades origin and three more
    objects that have an uncertain membership status due to their unknown or
    imprecise atmospheric parameters. Among those, two to three are expected
    as field stars contamination.  Accurate radial velocity measurements will
    confirm or reject the candidates.  One consequence is that the
    longstanding problem that no white dwarf older than $\sim$340 Myr appears
    to be associated with the cluster remains unsolved.}{}{}{}{}


   \titlerunning{Spectroscopic and photometric studies of white dwarfs in the Hyades}
   \authorrunning{Tremblay et al.}
   \maketitle

\section{Introduction}

The Hyades cluster, with a center-of-mass distance of 46.45 $\pm$ 0.5 pc
\citep{vanleeuwen09} and an estimated age of 625 $\pm$ 50 Myr, derived from
the helium abundance and isochrone modeling \citep{age}, is certainly one of
the most studied open clusters. Recently, \citet[][hereafter Paper
  I]{roeser11} identified 724 probable main-sequence members within a radius
of 30 pc from the cluster center. Their sample is based on a kinematic
selection from the PPMXL Catalog \citep{roeser10} with the convergent point
method, further constrained by photometric data. They determined individual
masses for their candidates and, among others, studied the mass distribution
as a function of the distance to the center. It was confirmed that the present
day tidal radius of the cluster is $\sim$9 pc, and that this value would only
change by a few percent under different assumptions on the fraction of
binaries, a quantity that is currently not well constrained.

In this work, we restrict our study to Hyades white dwarfs, even for which the
literature is extensive. It was already concluded long ago that old white
dwarfs with an age greater than $\sim$300 Myr were missing from the cluster
\citep{tinsley74,heuvel75}, and that they might have evaporated
\citep{pels75,weidemann77}. \citet{weidemann92} and \citet{eggen93} further
constrained this scenario by suggesting that old evaporated Hyades white
dwarfs form long tidal tails for which the density is much lower than that of
field stars. To this day, it is still the case that only 10 constituents are
confirmed, and N-body simulations systematically predict that at least two
times more white dwarfs were formed in the Hyades \citep{simul01,ernst11}. The
idea that many remnants evaporated from the cluster remains a likely scenario,
which is further suggested by the fact that Paper I found prominent tidal
tails for main-sequence stars, with almost as many members currently leaving
the cluster than members bound to the cluster. However, it is also possible
that the missing white dwarfs were simply never formed because of a very steep
initial mass function \citep{bc07} or that they are hidden in binaries
\citep{simul01}. \citet{williams04} tried to bring the observed number of
white dwarfs in the Hyades into agreement with calculated numbers by varying
the slope of the IMF, the binary fraction, the binary mass ratio and the
effect of dynamical evolution. He concludes that, from simulations alone, it
is not possible to differentiate between the above scenarios. A deficiency of
white dwarfs has also been observed in several other open clusters \citep[see,
  e.g.,][]{richer98,kalirai01,williams07}, although there are also many
instances where the observed and predicted numbers of degenerate objects are
in agreement \citep{kalirai01a,kalirai03,rubin08}.

In a recent analysis, \citet[][hereafter Paper II]{schilbach12} extended the
systematic search of Hyades members to white dwarfs up to 40 pc from the
cluster center. The authors did recover all 10 classical components, and
further identified 17 white dwarfs as new candidates, all of them outside of
the tidal radius. If these degenerate stars are confirmed as members, they
could potentially explain the discrepancy between the observed and predicted
number of associated objects, and also confirm that white dwarfs are slowly
leaving the cluster.

The 10 classical Hyades white dwarfs are certainly some of the most well known
and observed hot remnants. Given that new atmospheric parameters have recently
been published for most of these objects based on spectroscopic analyses
\citep{gianninas11,bergeron11} relying on improved model atmospheres
\citep{TB09}, it is certainly pertinent to review their properties as Hyades
representatives. Furthermore, most of the 17 new candidates identified in
Paper II also have well determined atmospheric parameters, from spectroscopic
or photometric analyses. Consequently, the next step to constrain their
membership is to compare quantities derived from kinematic studies,
e.g. kinematic distances, with those derived from their atmospheric
properties, which includes the so-called spectroscopic distance, based on
predicted luminosities \citep{holberg08}. The spectroscopic distances together
with the proper motions can be used to update the space velocities for
candidates, i.e. the residual 2D velocity with respect to the Hyades bulk
motion, which is important to understand the structure of the cluster.

This work is structured as follow. In Sect. 2, we describe the Hyades white
dwarfs sample of Paper II and Sect. 3 follows with a description of analyses
of their spectra and photometry. Based on their derived atmospheric
parameters, we compute, in Sect. 4, spectroscopic distances for all aspirant
Hyades remnants. In Sect.  5, we derive improved residual velocities, and try
to understand the global status of new candidates by simulating the field
white dwarfs contamination. Then, Sect. 6 continues with a discussion on the
membership of new candidates and the implications on the Hyades evolution. The
summary is presented in Sect. 7.

\section{Kinematic and photometric Hyades candidates}

The starting point of our analysis are the Hyades white dwarf candidates
identified in Paper II. The selection process is explained in details in Paper
I and II. In brief, the primary data of observations are the proper motions
and positions in the PPMXL catalog. The PPMXL is an all-sky survey with a
limiting magnitude of $V \sim 20$ including about 900 million objects. The
PPMXL data was cross-matched and combined with the UCAC3 \citep{zacharias10}
and the CMC14 \citep{CMC14}, and referred as a whole as the Carlsberg-UCAC
(CU) subset. Along with the kinematics, this data set also includes 2MASS
$JHK_S$ and CMC14 $r'$ magnitudes.

Candidates are first selected using the convergent point method (Paper
I). More precisely, in the case of white dwarf candidates, an upper limit of 5
km~s$^{-1}$ is allowed for the difference between their tangential motion and
the bulk motion of the Hyades. The convergent point method then predicts, for
each possible member, a radial velocity and a distance, hereafter named the
kinematic distance ($d_{\rm kin}$), based only on the position ($\alpha$,
$\delta$) and proper motion ($\mu_\alpha$, $\mu_\delta$). To avoid
contamination by field objects, only the new candidates with a $d_{\rm kin}$
prediction that put them closer than 40 pc from the cluster center are kept.

The second step used in Paper II to construct the Hyades sample was to
cross-match all kinematically suited candidates with white dwarf catalogues,
such as the online version of the Villanova White Dwarf
Catalog\footnote{http://www.astronomy.villanova.edu/WDCatalog/index.html}
\citep{mccook99}. From this procedure, the authors identified a total of 37
white dwarfs as kinematic candidates. We note that at this point, a fair
number of these objects could be background objects with proper motions that
mimic the Hyades motion.  Therefore, the final step was to examine the
available photometric data of the candidates. These white dwarfs are
unfortunately at the detection limit of the 2MASS and CMC14 survey. Even if
$JHK_s$ and $r'$ magnitudes are available for most candidates, a high scatter
was found in the $J-K_S$ colors, which is consistent with the conclusion of
\citet{TB07} that one should be cautious with $K_s$ magnitudes near the faint
detection limit of the survey. Therefore, it was opted to include in the
analysis $B$ and $V$ magnitudes from the literature even if the accuracy of
the available $BV$ was far from being uniform and varied from several millimag
(photoelectric measurements) to several tenths of a mag (photographic
measurements).

\begin{figure}[h!]
\begin{center}
\includegraphics[bb=10 200 592 682 ,width=3.5in]{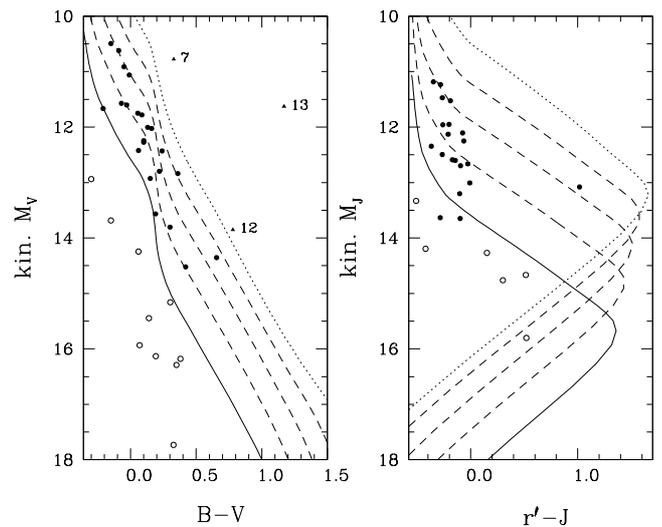}
\caption{Kinematic absolute magnitude versus color diagrams for Hyades white
  dwarf candidates (filled circles) and probable non-Hyades members (open
  circles) identified in Paper II. Binaries are marked as filled triangles and
  by their numbers from Table~1, although in most cases, they are too bright
  for the range of absolute magnitudes shown in the figure. Theoretical
  cooling sequences are shown from $\log g = 7.0$ (dotted line) to $\log g =
  9.0$ (solid line) in steps of 0.5 dex.
\label{fg:f1}}
\end{center}
\end{figure}

Using the kinematic distance and an observed magnitude, a ``kinematic''
absolute magnitude (kin. $M$) can be computed. In Fig.~\ref{fg:f1} we
reproduce the kin. $M_V$ vs. $B-V$ and kin. $M_J$ vs. $r'-J$ diagrams made in
Paper II, but now compared with theoretical white dwarf cooling sequences
\citep{wood95,fontaine01}. True Hyades members are expected to follow the
theoretical sequences. On the contrary, background and foreground stars are
expected to be shifted vertically in relation to the theoretical sequences,
because their estimated $d_{\rm kin}$ differs from the true distance.  We can
see from Fig.~\ref{fg:f1} that the 10 white dwarfs identified as probable
non-members in Paper II (open circles) lie below the theoretical cooling
sequences. They would have to be super massive ($\log g >$ 9.0) to be cluster
members. Instead, they are much more likely distant objects mimicking the
Hyades proper motion.

All 10 of the so-called classical Hyades members were recovered in Paper
II. Removing 10 white dwarfs identified as probable non-members, as discussed
above, there remain 17 new candidates, all lying outside of the tidal radius
of $\sim$9 pc. However, five of the remaining candidates are more than 20 pc
away from the cluster center in the $Z$ direction (perpendicular to the
galactic plane) and main-sequence stars with known radial velocities in these
regions have been ruled out as members (Paper I). We note that three of the
classical members and two of the new candidates are unresolved binaries with a
main-sequence companion.

\subsection{New selection of Hyades candidates}

The kinematic distances of the Hyades white dwarf candidates were obtained in
Paper II by the requirement of the minimum difference between their tangential
velocity and the cluster motion, with an upper limit of the residual velocity
of 5 km~s$^{-1}$. Later in this work (see Sect. 5 below) we recalculate
residual velocities of the candidates by assuming the spectroscopic distances
derived in Sect.~4. We found that most of the candidates outside of the tidal
radius have residual velocities up to 8 km~s$^{-1}$. Therefore, we conclude
that an upper limit of 5 km~s$^{-1}$ for the residual velocity may have been a
too rigorous requirement for detecting escaping white dwarfs.

To obtain a complete sample of possible Hyades candidates, we revisited the
selection process of Paper II by using a higher upper limit of 8 km~s$^{-1}$
for the residual velocity. We found 14 additional white dwarfs for which the
kinematic requirement was fulfilled. All but three candidates were later
rejected by spectroscopic criteria. The three promising candidates are listed
as star Nos. 38 to 40 at the bottom of Table~1.

\section{Spectroscopic and photometric stellar parameters}

An improved diagnostic of the Hyades membership of white dwarfs can be made
based on known stellar parameters. In Table~1, we have compiled the $T_{\rm
  eff}$, $\log g$ and total stellar masses found in the literature for 37 of
the 40 candidates. The corresponding absolute V magnitudes ($M_V$) and white
dwarf cooling time ($\log \tau$) have then been computed using the
evolutionary models with thick layers (hydrogen-rich atmosphere DA white
dwarfs) and thin layers (helium-rich atmosphere DB white dwarfs) of
\citet[][carbon-oxygen cores]{fontaine01} for $T_{\rm eff} < 30,000$~K and
\citet[][carbon cores]{wood95} above this temperature. When available in the
cited literature, we also compiled the distances, although we note that the
calculation of improved distances is the goal of our Sect. 4 below. Omitted
from Table~1 is the binary white dwarf WD~0217+375 (No. 13), which was
designated as an M5V+DA pair without a detailed spectral analysis of the white
dwarf \citep{silvestri05}. Also omitted is LP~649-0071 (No. 12), a M dwarf
star with a suspected but spectroscopically unconfirmed white dwarf companion,
previously identified in Luyten's White Dwarf Catalogues
\citep{luyten99}. Finally, the unconfirmed white dwarf and probable non-member
1RXSJ062052.2+132436 (No. 34) also has no precise atmospheric parameters.

The majority of the candidates are hydrogen-rich DA white dwarfs. The most
successful and precise technique to identify the atmospheric parameters of DA
white dwarfs is to fit the observed Balmer line profiles with synthetic
profiles predicted from model atmospheres, which is called the spectroscopic
technique \citep{weidemann80,schulz81,bergeron92}. As much as 26 objects in
Table~1 rely on the spectroscopic analysis of the Balmer lines to identify
their atmospheric parameters (references 2, 7, 11 and 12 in the table). Among
them, 22 are from the White Dwarf Catalog survey of \citet{gianninas11} and
two are from the updated SDSS analysis of \citet{TB11}. Both of these analyses
adopted the most recent model atmospheres of \citet{TB09}, relying on Balmer
line profiles including non-ideal gas effects. The quoted uncertainties in
Table~1 are from the fitting procedure. However, more realistic estimates of
the uncertainties can be made by fitting different observations of the same
objects, and \citet{LBH05} found spreads of 1.2$\%$ and 0.038 dex in $T_{\rm
  eff}$ and $\log g$, respectively. For the remaining of this work, we use
these values as the {\it minimum} uncertainties on the atmospheric
parameters. We note that possible systematic uncertainties from the models are
not included. Yet, promising Hyades candidates (Nos. 1-27) with spectroscopic
parameters in Table~1 have radiative atmospheres and are outside the range of
the so-called high-mass problem \citep{koester09a,tremblay11}. For these
objects ($T_{\rm eff} > 13,000$ K), the atmospheric parameters are in good
agreement with the few observational constraints, such as parallax
measurements \citep[see, e.g., Fig. 14 of][]{TB09}. We emphasize that the
current work can also serve as a validation for recent DA models since the
distance to the Hyades center is fairly well constrained via Hipparcos
\citep{vanleeuwen09}. Finally, the helium and hydrogen lines in the DBA star
WD~0437+138 have been fitted spectroscopically by \citet{bergeron11} with
improved helium-rich models.

Among the remaining candidates, WD~0433+270 (No. 21) is a cool DA with only a
faint H$\alpha$ line for which a spectroscopic fit is not possible. The
$BVRIJHK$ photometric energy distribution of this object has been fitted by
\citet{bergeron01} to determine $T_{\rm eff}$ and also the $\log g$ using the
constraint from the trigonometric parallax. We use here the results from
\citet{giammichele12} who refitted the data using models with the
Lyman~$\alpha$ H$-$H$_2$ neutral broadening \citep{kowalski06}. The best fit
parameters were used to predict a H$\alpha$ line profile that is in close
agreement with the observed one (see Fig. 11f from Giammichele et al.), and
therefore we can consider that the atmospheric parameters for this object are
similarly precise as those found with the spectroscopic method for the
previously discussed stars. The cool and nearby magnetic DQP WD~0548$-$001
(No. 39) has also been fitted by \citet{giammichele12} using a similar method,
although with an additional free parameter due to the presence of carbon in
the helium-rich atmosphere. All in all, 29 of the 37 objects in Table~1 are
typical DA, DB or DQ with well constrained stellar parameters. We remind the
reader, however, that two magnetic objects (Nos. 24 and 39) and two objects
that have been fitted without the improved Stark broadening profiles (Nos. 20
and 35) likely have underestimated uncertainties.

Other objects in Table~1 have more uncertain atmospheric
parameters. WD~0816+376 (No. 27) is a DAP white dwarf with a high magnetic
field that is part of the Kiso survey sample of \citet{limoges10}. Even if the
object is included in their final sample, they used the estimation of
\citet{jordan93} for the temperature and fixed the gravity at $\log g$ =
8.0. Therefore, we chose a characteristic uncertainty of 0.5 dex for the
gravity of this object. The situation is similar for the poorly studied cool
DC remnant WD~0648+368 (No. 40) and the DQA SDSS~J023614.44$-$080804.9
(probable non-member) for which we also fixed the gravity.

Of the 4 objects in Table~1 that are unresolved binaries (Nos. 1, 4, 7 and
36), WD~0429+176 and WD~0816+387 can be fitted relatively well using the
spectroscopic method by removing iteratively the M dwarf contribution. In
contrast, the classical remnants WD~0347+171 and WD~0418+137 have F- and
K-type main-sequence companions that completely dominate the optical flux. As
a consequence, these two white dwarfs were observed in the UV to constrain
their atmospheric parameters, and the spectroscopic method applied to the
Lyman lines is not as well documented and precise as for the Balmer line
analysis \citep{vennes05}. For one of them, the gravity could not be
constrained from the Lyman~$\alpha$ spectrum alone, and we chose a
characteristic value of $\log g = 8.0 \pm 0.5$.

We have at hand, from the PPMXL and paper II, a photometric data set with
Johnson $BV$, CMC14 $r'$ and 2MASS $JHK_S$ magnitudes for most objects in
Table~1. We therefore proceeded to fit the photometric data with grids of
pure-H and pure-He white dwarf models, with $T_{\rm eff}$ and the solid angle
$R/D$, where $R$ is the stellar radius and $D$ the distance from the Sun, as
free parameters. This method is not sensitive to the gravity, and we used a
fixed value of $\log g = 8.0$. Consequently, this method can only be used to
give an independent value\footnote{Note that the distances could also be
  determined with this method, but the technique described in Sect. 4 is more
  precise.}  of $T_{\rm eff}$, called photometric temperature below.  We rely
on the prescription of \citet{holberg06} to convert the magnitudes into
average fluxes.

\begin{figure}[h!]
\begin{center}
\includegraphics[bb=70 400 302 692,width=2.5in]{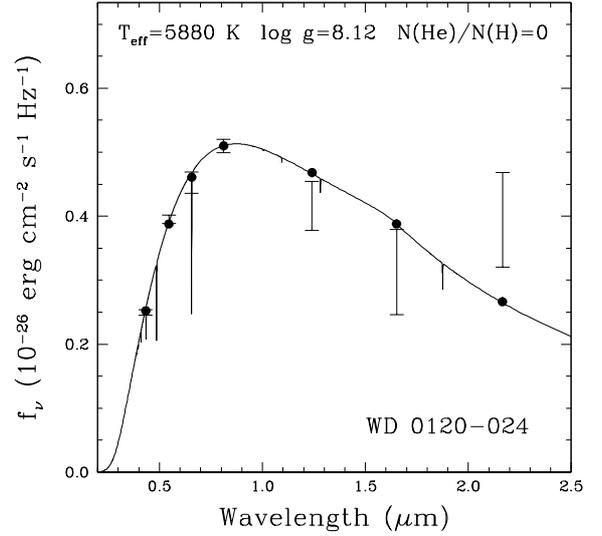}
\caption{Johnson $BVRI$ and 2MASS $JHK_S$ photometry of WD~0120$-$024 (error
  bars) fitted by average model fluxes (filled points) for a pure-hydrogen
  atmosphere. The best fit parameters are given in the figure, and the
  corresponding predicted spectrum is also shown (solid line).
\label{fg:f2}}
\end{center}
\end{figure}

From our photometric fits, we find that in all but one case the photometric
temperature agrees with the spectroscopic value. The exception is
WD~0231$-$054 for which the $JHK$ flux appears too high and suggests a lower
$T_{\rm eff}$ for this object. We have no clear explanation for this
behaviour. Most objects in Table~1 are hot and blue, and their colors, in the
Rayleigh-Jeans regime, are not very sensitive to $T_{\rm eff}$ (or the
absolute magnitude, as can be seen in Fig.~\ref{fg:f1}). Therefore, the
photometric temperatures are less precise and we rely, whenever possible, on
spectroscopically derived temperatures.  There are four cases where no
spectroscopic parameters are available. The first is the DA WD~0120$-$024
(No. 11), which has only been fitted using the $V-I$ color in
\citet{silvestri01}. On the contrary, our data set includes 2MASS $JHK_S$
photometry, and we could also find for this object the Johnson $BVRI$
photometry \citep{smith97}.  Our fit is shown in Fig.~\ref{fg:f2} with the
gravity fixed at the value found by \citet{silvestri01} from a gravitational
redshift measurement. We note a discrepancy at the $K_S$ band, which is not
unusual as it is close to the detection limit of 2MASS. The atmospheric
parameters should be considered as relatively well constrained for this
object, despite the fact that masses determined from gravitational redshifts
are generally not as precise as those found from spectroscopic or parallax
measurements \citep{bergeron07}.

\begin{sidewaystable*}
\vspace*{18cm}
 \caption{Atmospheric parameters of Hyades white dwarf candidates}
 \label{tab1}
 \begin{tabular}{llllcccccccl}
\hline
\hline
Star No. & WD name & Other name & SpT & $T_{\rm eff}$ & $\log g$ & $M/M_{\odot}$ & M$_V$ & $\tau$ & dist & Ref.\tablefootmark{a}\\
         &         &            &     &      (K)      & (cm s$^{-2}$)      &               &  (mag)  & (yr)    &   (pc) &   \\
\hline

 1 & 0347+171     & V471 Tau                   & DA+dK  &  34100 (100) & 8.25 (0.05) & 0.84 (0.04) & 10.20 (0.12) & 7.07 (0.13) &  $-$ & 1  \\
 2 & 0352+096     & HZ 4                       & DA     &  14670 (380) & 8.30 (0.05) & 0.80 (0.03) & 11.76 (0.09) & 8.53 (0.05) &   36 & 2  \\
 3 & 0406+169     & LB 227                     & DA     &  15810 (290) & 8.38 (0.05) & 0.85 (0.03) & 11.76 (0.09) & 8.50 (0.04) &   52 & 2  \\
 4 & 0418+137     & HD 27483                   & DA+dF  &  21410 (210) & 8.00 (0.50) & 0.63 & 10.65 & 7.70 &  $-$ & 3  \\
 5 & 0421+162     & LP 415-46                  & DA     &  20010 (320) & 8.13 (0.05) & 0.70 (0.03) & 10.96 (0.08) & 7.97 (0.06) &   47 & 2  \\
 6 & 0425+168     & LP 415-415                 & DA     &  25130 (380) & 8.12 (0.05) & 0.71 (0.03) & 10.53 (0.08) & 7.49 (0.08) &   50 & 2  \\
 7 & 0429+176     & HZ 9                       & DA+dM  &  17620 (350) & 8.02 (0.06) & 0.63 (0.04) & 11.02 (0.09) & 8.08 (0.06) &   39 & 2A \\
 8 & 0431+126     & HZ 7                       & DA     &  21890 (350) & 8.11 (0.05) & 0.69 (0.03) & 10.78 (0.08) & 7.78 (0.07) &   49 & 2  \\
 9 & 0437+138     & LP 475-242                 & DBA    &  15120 (360) & 8.25 (0.09) & 0.74 (0.06) & 11.65 (0.15) & 8.47 (0.07) &   45 & 4  \\
10 & 0438+108     & HZ 14                      & DA     &  27540 (400) & 8.15 (0.05) & 0.73 (0.03) & 10.40 (0.08) & 7.30 (0.09) &   49 & 2  \\
\hline
11 & 0120$-$024   & LP 587-53                  & DA     &   5880 (80)  & 8.12 (0.16) & 0.66 (0.10) & 14.53 (0.24) & 9.51 (0.15) &  $-$ & 5  \\
14 & 0230+343     & GD 30                      & DA     &  15270 (250) & 7.98 (0.05) & 0.60 (0.03) & 11.21 (0.08) & 8.26 (0.04) &   91 & 2  \\
15 & $-$          & LP 246-14                  & $-$    &   9680 (2600)& 8.00 (0.50) & 0.60 & 12.31 & 8.82 &  $-$ & 6  \\
16 & 0259+378     & GD 38                      & DA     &  32040 (470) & 7.84 (0.05) & 0.57 (0.03) &  9.59 (0.09) & 6.89 (0.02) &  174 & 2  \\
17 & 0312+220     & GD 43                      & DA     &  18600 (300) & 7.89 (0.05) & 0.56 (0.03) & 10.74 (0.08) & 7.86 (0.06) &  105 & 2  \\
18 & 0339$-$035   & LP 653-26                  & DA     &  13000 (210) & 8.06 (0.05) & 0.65 (0.03) & 11.60 (0.07) & 8.51 (0.04) &   52 & 2  \\
19 & 0348+339     & GD 52                      & DA     &  14820 (350) & 8.31 (0.05) & 0.80 (0.03) & 11.76 (0.09) & 8.52 (0.05) &   49 & 2  \\
20 & $-$          & HS 0400+1451               & DA     &  14620 (60)  & 8.25 (0.01) & 0.76 (0.01) & 11.69 (0.02) & 8.50 (0.01) &  $-$ & 7  \\
21 & 0433+270     & LP 358-525                 & DA     &   5630 (100) & 8.06 (0.04) & 0.62 (0.02) & 14.65 (0.10) & 9.51 (0.06) &   18 & 8  \\
22 & 0437+122     & LP 475-249                 & DA     &  13260 (350) & 8.08 (0.08) & 0.66 (0.05) & 11.60 (0.12) & 8.50 (0.06) &  165 & 2  \\
23 & 0625+415     & GD 74                      & DA     &  17610 (280) & 8.07 (0.05) & 0.66 (0.03) & 11.10 (0.08) & 8.12 (0.05) &   60 & 2  \\
24 & 0637+477     & GD 77                      & DAP    &  14650 (590) & 8.30 (0.06) & 0.80 (0.04) & 11.76 (0.12) & 8.53 (0.06) &   40 & 2B \\
25 & 0641+438     & LP 205-27                  & DA     &  16400 (270) & 8.00 (0.05) & 0.62 (0.03) & 11.12 (0.08) & 8.17 (0.05) &   76 & 2  \\
26 & 0743+442     & GD 89                      & DA     &  15220 (350) & 8.47 (0.05) & 0.91 (0.03) & 11.98 (0.09) & 8.61 (0.05) &   39 & 2  \\
27 & 0816+376     & GD 90                      & DAP    &  11000 (630) & 8.00 (0.50) & 0.60 & 11.87 & 8.67 &  $-$ & 9  \\
\hline
28 & 0233$-$083.1 & SDSS J023614.44$-$080804.9 & DQA    &  10000 (1000) & 8.00 (0.50) & 0.60 & 12.18 & 8.78 &  $-$ & 10 \\
29 & 0300$-$083.1 & SDSS J030325.22$-$080834.9 & DA     &  11550 (200) & 8.55 (0.08) & 0.96 (0.05) & 12.62 (0.15) & 9.00 (0.09) &  $-$ & 11 \\
30 & $-$          & LP 652-342                 & $-$    &  10940 (2840) & 8.00 (0.50) & 0.60 & 11.88 & 8.68 & $-$ & 6 \\
31 & 0533+322     & G98-18                     & DA     &  16970 (280) & 7.59 (0.05) & 0.43 (0.02) & 10.47 (0.08) & 8.09 (0.03) &  159 & 2  \\
32 & 0543+436     & G96-53                     & DAZ    &   8260 (130) & 7.46 (0.13) & 0.36 (0.05) & 12.19 (0.19) & 9.03 (0.07) &   99 & 2  \\
33 & 0557+237     & G104-10                    & DA     &   8360 (120) & 8.18 (0.06) & 0.70 (0.04) & 13.15 (0.11) & 9.10 (0.05) &   56 & 2  \\
35 & 0758+208     & SDSS J080114.02+204340.3   & DA     &   7640 (70) & 8.68 (0.20) & 1.03 (0.12) & 14.35 (0.38) & 9.56 (0.04) &  $-$ & 12 \\
36 & 0816+387     & G111-71                    & DA+dM  &   7700 (110) & 8.07 (0.07) & 0.64 (0.04) & 13.31 (0.12) & 9.12 (0.05) &   45 & 2A \\
37 & 0820+250.1   & SDSS J082346.14+245345.6   & DA     &  34050 (100) & 7.74 (0.02) & 0.53 (0.01) &  9.29 (0.03) & 6.80 (0.01) &  $-$ & 11 \\
\hline
38 & 0231$-$054   & PHL 1358                   & DA     &  17470 (270) & 8.58 (0.05) & 0.98 (0.03) & 11.94 (0.09) & 8.52 (0.04) &   29 & 2  \\
39 & 0548$-$001   & G99-37                     & DQP    &   6130 (100) & 8.18 (0.04) & 0.69 (0.03) & 14.42 (0.10) & 9.56 (0.05) &   11 & 8  \\
40 & 0648+368     & GD 78                      & DC     &   5700 (1000) & 8.00 (0.50) & 0.59 & 14.51 & 9.42 &   $-$ & 13  \\

\end{tabular} 
\tablefoot{Star Nos. and categories are the same as in Paper II. Nos. 1 to 10
  are the so-called classical Hyades white dwarfs. Nos. 11 to 27 are new
  candidates (including star Nos. 23 to 27 that are more than 20 pc away from
  the center in the Z direction). Nos. 28 to 37 are rated probable non-Hyades
  members. Nos. 38 to 40 are new candidates identified in Sect. 2
  with a residual velocity upper limit of 8 km s$^{-1}$.}
\tablebib{ (1) \citet{sion12}; (2) \citet{gianninas11};
  (2A) same as previous but with a M-dwarf component removed; (2B) same as
  previous but magnetic white dwarf fitted with non-magnetic models; (3)
  \citet{burleigh98}; (4) \citet{bergeron11}; (5) photometric fit for $T_{\rm
    eff}$, this work, and mass from \citet{silvestri01}; (6) photometric fit,
  this work; (7) \citet{koester09b}; (8) \citet{giammichele12}; (9)
  \citet{limoges10,jordan93}; (10) \citet{SDSSmag}; (11) \citet{TB11}; (12)
  \citet{SDSS}; (13) \citet{angel81}.}
\end{sidewaystable*}

\begin{figure}[h!]
\begin{center}
\includegraphics[bb=90 400 512 702,width=2.7in]{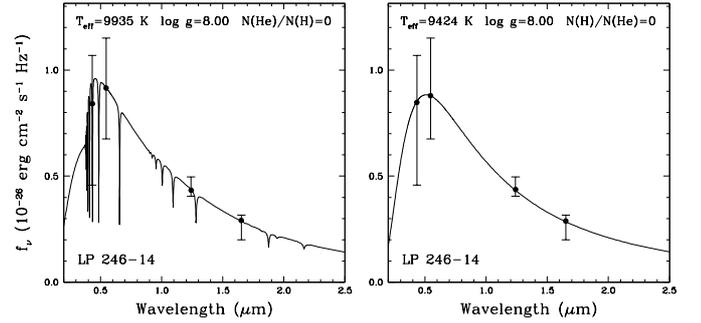}
\caption{Similar to Fig.~\ref {fg:f2} but for the $BV$ and 2MASS $JH$ fit of
  LP~246-14 using pure-hydrogen models (left panel) and pure-helium models
  (right panel).
\label{fg:f3}}
\end{center}
\end{figure}

In Fig.~\ref{fg:f3} we show our $BV$ and 2MASS $JH$ fit of LP~246-14 (No. 15),
which is found in the Luyten's White Dwarf Catalogues but is not a
spectroscopically confirmed white dwarf, and no atmospheric parameters have
yet been published. The fit is consistent with a relatively cool white dwarf,
but the lack of a spectrum for this object prevents us to conclude that it is
a white dwarf. While in the next section it is shown that the $d_{\rm kin}$
value is consistent with a gravity of $\log g \sim 8$, it does not exclude
that it is a background lower gravity object with a high tangential
velocity. Therefore, the atmospheric parameters presented in Table~1, an
average of the pure-H and pure-He solutions, should be used with high caution
until a spectrum is secured. Among the other possible but unconfirmed white
dwarfs, we derived the atmospheric parameters of LP~652-342 (probable
non-member) in a similar fashion. For 1RXSJ062052.2+132436, we find that the
photometry is consistent with a very hot object, but it is in a regime where
$T_{\rm eff}$ can not be estimated with a reasonable accuracy, and therefore
it is omitted from Table~1.

\section{Spectroscopic and Kinematic distances}

\begin{figure*}[]
\begin{center}
\includegraphics[bb=90 210 582 652,width=4.4in]{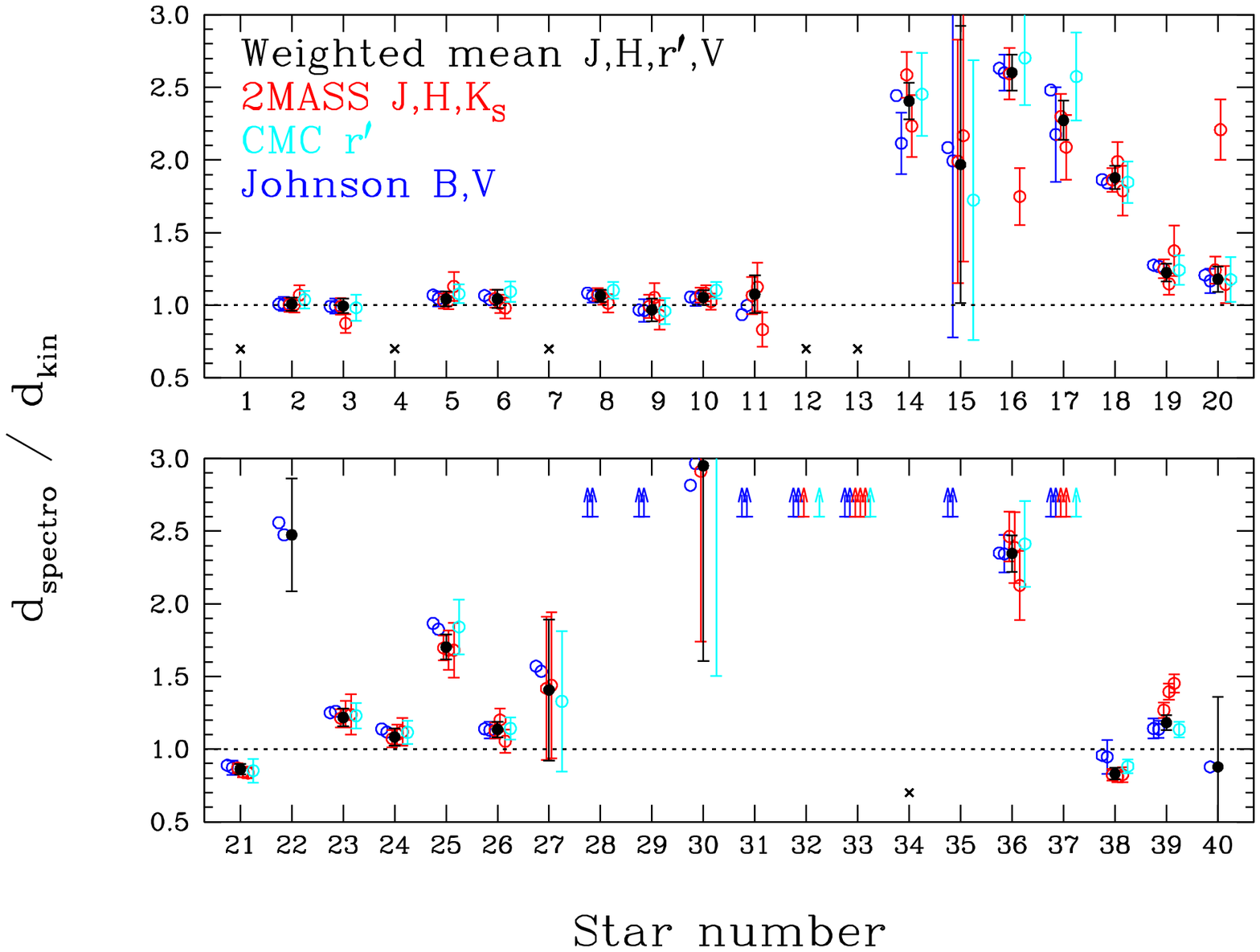}
\caption{Ratio of the spectroscopic to kinematic distance for the Hyades white
  dwarf candidates identified by their star number from Table~1 (from No. 1 to
  20 in upper panel and from No. 21 to 40 in the lower panel). Up to five
  $d_{\rm spectro}$ values were evaluated for each candidate using 2MASS
  $JHK_S$ (red), CMC $r'$ (cyan) and Johnson $BV$ (blue) photometry (open
  points). Many $B$ and $V$ magnitudes have no published uncertainties. Also
  shown is the the mean weighted spectroscopic distance from 2MASS $JH$, CMC
  $r'$ and $V$ filters (black filled points). The data points are shifted
  horizontally for clarity. Objects with a cross above the star No. are
  binaries (Nos. 1, 4, 7, 12 and 13) or have no atmospheric parameters
  (No. 34). For several objects, the ratio is actually higher than the scale
  of the plot, and is represented by a lower limit symbol. The dotted
  horizontal line represents a perfect match between both distance estimates.
\label{fg:f4}}
\end{center}
\end{figure*}

The set of atmospheric parameters that we have compiled in Table~1 can be
used, along with model atmospheres, to predict absolute magnitudes for any
photometric band.  By combining these values with observed magnitudes, we can
compute the so-called {\it spectroscopic} distances ($d_{\rm
  spectro}$)\footnote{Note that in a few cases, photometric instead of
  spectroscopic fits were used to determine the atmospheric parameters, but we
  use the term spectroscopic distance in this work for clarity.}. This is
independent of the {\it kinematic} distances derived in Paper II to which they
can be compared.  Our photometric data set, including 2MASS $JHK_S$, CMC14
$r'$ and Johnson $BV$ can be used to predict as much as 6 alternative $d_{\rm
  spectro}$ values. In Fig.~\ref{fg:f4} we present, for all observed
magnitudes, the ratio of the spectroscopic and kinematic distances for objects
in Table~1. In the computation of the uncertainties, we account for the
contribution of the $d_{\rm kin}$ prediction (depending only on the observed
proper motion and sky position), the atmospheric parameters and the observed
magnitude. We note that unresolved binaries are omitted from Fig.~\ref{fg:f4},
since the observed magnitudes are all either mildly or strongly contaminated
by the companion.

It is clear that the different $d_{\rm spectro}$ determinations for a same
object are not completely independent, since they are based on the same set of
atmospheric parameters. Nevertheless, by looking at the results between
different filters, one can get an overview of the quality of the photometric
data sets. It is seen that the scatter is in general low between the different
filters, implying that our photometric data sets are fairly homogeneous and
well calibrated. We nevertheless note that one $K_S$ magnitude (No. 16) and
one $H$ magnitude (No. 20) in Fig.~\ref{fg:f4} are significantly different to
other values. \citet{TB07} have shown that one must be careful with spurious
magnitude values in 2MASS close to the survey detection limit, and these are
unlikely to be the signature of real near-infrared spectral
features. Furthermore, $B$ and $V$ magnitudes are drawn from different sources
with inhomogeneous accuracy, and often even have no quoted
uncertainties\footnote{Our results indicate that $B$ and $V$ magnitudes are
  nevertheless in fairly good agreement with the other data sets, and that the
  often used typical average uncertainty of 0.05 mag in $V$ \citep[see,
    e.g.,][]{TB08} appears adequate.}.  Based on these facts, we decided to
compute a weighted mean $d_{\rm spectro}$ value using only 2MASS $JH$, CMC
$r'$ and $V$ magnitudes (when the uncertainties are given)\footnote{For stars
  No. 22 and 40, we assume a V uncertainty of 0.2 since no 2MASS or CMC
  measurements are available.}. In Fig.~\ref{fg:f4}, we show the comparison
between the mean $d_{\rm spectro}$ and $d_{\rm kin}$ determinations for each
candidate. These quantities are also given in Table~2, along with the number
of filters involved in the $d_{\rm spectro}$ mean. Note that at this point, we
leave out from our calculations and discussion the objects labelled as
probable non-members in Paper II (Nos. 28-37), since they are clearly all
background objects according to Fig.~\ref{fg:f4}.

In the next section we will discuss about the Hyades membership of new
candidates based on the results of Fig.~\ref{fg:f4}. Beforehand, there is some
interesting information that can be extracted from the classical members. The
non-binary Hyades white dwarfs (Nos. 2-3, 5-6 and 8-10) can serve as a
benchmark to evaluate the precision of our analysis. The spread due to the
photometric uncertainties has already been evaluated from Fig.~\ref{fg:f4}. It
is now seen from the same figure that the mean $d_{\rm spectro}$ values
derived in this work are in good agreement, roughly within the uncertainties,
with $d_{\rm kin}$ values. The spread around the expected unity ratio can be
used to learn about the combined uncertainties due to spectroscopic parameter
determinations and $d_{\rm kin}$ predictions. The good agreement suggests that
the uncertainties in this work are relatively well estimated and that our
method should be fairly robust to argue on the status of the Hyades membership
of new candidates. Furthermore, we note that the single classical members have
a mean mass of $0.75$ $M_{\odot}$, significantly higher than the value of
$0.638$ $M_{\odot}$, with a dispersion of 0.145 $M_{\odot}$, found for field
white dwarfs \citep{gianninas11}.  Also, their cooling ages are in the range
10-340 Myr, which is consistent with the young age of the Hyades, though
significantly smaller than the estimated age of the cluster of 625
Myr. According to \citet{dobbie06}, this corresponds to main-sequence
progenitors in the range of $\sim$2.5-4 M$_{\odot}$. Therefore, more massive
and slightly older white dwarfs, hence fainter and cooler objects, could be
expected among the new candidates.

One uncertainty that is often difficult to evaluate, and not included in
Fig.~\ref{fg:f4}, is the precision of the model atmospheres. The classical
Hyades white dwarfs, however, can help us for the evaluation of this
uncertainty. Since any possible inaccuracies in the model atmospheres are
expected to cause a systematic offset of the atmospheric parameters, they
would also cause systematic shifts of the points in Fig \ref{fg:f4}. Other
uncertainties discussed so far are much more likely to cause a spread around a
mean value. Our results show that there is no need to account for an
additional offset, and therefore we conclude that the atmospheric parameters
are sufficiently accurate on the average. This is a fairly precise validity
verification of the physics in the \citet{TB09} DA models, and to a lesser
degree the \citet{bergeron11} DB models (since only star No. 9 is
helium-rich), which were adopted to predict the atmospheric parameters of
classical members. Reversing this statement, it illustrates that the method of
using model atmospheres and spectroscopic observations to predict $d_{\rm
  spectro}$ values, which has become a standard method in recent years
\citep[see, e.g.,][]{holberg08,dobbie09,williams09}, is fairly well calibrated
and precise on average. One still has to be cautious about the precision of
observed magnitudes used in this process.

 \begin{table*}[]
 \begin{center}
 \caption{Distances of Hyades white dwarf candidates}
 \label{tab2}
 \begin{tabular}{llcccccl}
\hline 
\hline 
Star No. & Name & $d_{\rm kin}$ & $d_{\rm spectro}$ & Filters\tablefootmark{a} & $\vert v_{\rm tang} \vert$\tablefootmark{b} & $d_{\rm Hyades~center}$\tablefootmark{b} & Status\tablefootmark{c} \\ 
         &      &      (pc) &  (pc)  &   &  (km s$^{-1}$) & (pc) &  \\ 
\hline

 2 & WD 0352+096   &  35.1 (0.7) &  35.4 (1.4)  & 4 &   1.0 (0.9, 1.7) &  13.3 (12.2, 14.4) & classical \\
 3 & WD 0406+169   &  51.8 (1.6) &  51.5 (2.0)  & 4 &   0.5 (0.4, 1.2) &   6.5 ( 5.0,  8.2) & classical \\
 5 & WD 0421+162   &  44.1 (1.4) &  46.0 (1.7)  & 4 &   1.4 (0.9, 2.2) &   1.0 ( 0.9,  2.2) & classical \\
 6 & WD 0425+168   &  47.8 (2.1) &  49.9 (2.0)  & 3 &   1.2 (0.6, 2.2) &   3.6 ( 1.6,  5.6) & classical \\
 8 & WD 0431+126   &  45.7 (0.4) &  48.6 (1.9)  & 4 &   1.5 (0.6, 2.4) &   4.2 ( 3.4,  5.5) & classical \\
 9 & WD 0437+138   &  46.9 (1.8) &  45.4 (3.2)  & 4 &   1.0 (0.7, 2.2) &   3.3 ( 3.2,  5.2) & classical \\
10 & WD 0438+108   &  46.1 (1.1) &  48.6 (1.9)  & 4 &   1.2 (0.4, 2.0) &   5.8 ( 5.3,  6.9) & classical\\
11 & WD 0120$-$024 &  39.1 (0.8) &  42.0 (5.0)  & 2 &   3.5 (0.8, 9.1) &  37.1 (35.9, 39.0) & old WD \\
14 & WD 0230+343   &  37.9 (0.6) &  91.1 (4.5)  & 4 &  56.0 (51.3, 60.8) &  56.8 (52.8, 60.9) & non-member \\
15 & LP 246-14     &  35.1 (0.5) &  69.1 (33.4) & 4 &  37.9 (2.5, 75.0) &  36.8 (23.3, 66.3) & old WD \\
16 & WD 0259+378   &  66.2 (1.8) & 172.3 (6.9)  & 3 &  61.9 (57.9, 65.8) & 133.4 (126.6, 140.2) & non-member \\
17 & WD 0312+220   &  44.4 (1.0) & 101.0 (5.6)  & 4 &  45.3 (40.9, 49.7) &  58.7 (53.3, 64.1) & non-member \\
18 & WD 0339$-$035 &  28.5 (0.4) &  53.5 (2.1)  & 3 &  27.5 (25.4, 29.8) &  21.2 (20.2, 22.4) & non-member \\
19 & WD 0348+339   &  38.5 (0.7) &  47.1 (2.1)  & 3 &   7.3 (5.6, 9.1) &  15.6 (15.3, 16.2) & candidate \\
20 & HS 0400+1451  &  38.5 (2.5) &  45.2 (1.3)  & 3 &   5.1 (4.2, 6.0) &   5.1 ( 5.0,  5.4) & candidate \\
21 & WD 0433+270   &  20.1 (0.2) &  17.3 (0.7)  & 4 &   3.6 (2.7, 4.5) &  29.5 (28.8, 30.2) & old WD \\
22 & WD 0437+122   &  67.1 (7.7) & 166.1 (17.9) & 1 &  31.7 (26.0, 37.4) & 120.0 (102.2, 137.9) & non-member \\
23 & WD 0625+415   &  47.6 (1.4) &  58.0 (2.4)  & 3 &   6.6 (5.5, 7.9) &  34.0 (32.6, 35.5) & candidate \\
24 & WD 0637+477   &  36.2 (0.7) &  39.3 (2.1)  & 3 &   3.0 (1.7, 4.6) &  30.9 (30.6, 31.2) & candidate \\
25 & WD 0641+438   &  41.8 (1.1) &  71.2 (3.1)  & 3 &  19.9 (17.9, 21.9) &  46.2 (43.8, 48.6) & non-member \\ 
26 & WD 0743+442   &  33.6 (0.6) &  38.1 (1.7)  & 4 &   4.3 (2.9, 5.8) &  36.4 (36.0, 36.8) & candidate \\
27 & WD 0816+376   &  38.8 (0.9) &  54.5 (18.8) & 3 &  11.8 ( 0.9, 25.9) &  46.8 (38.8, 59.9) & uncertain\\ 
38 & WD 0231$-$054 &  32.8 (0.4) &  27.2 (1.3)  & 3 &   8.6 (7.6, 9.7) &  29.0 (28.6, 29.5) & candidate\\
39 & WD 0548$-$001 &  9.4 (0.2)  &  11.1 (0.4)  & 3 &   6.8 (6.4, 7.1) &  36.7 (36.4, 37.1) & old WD\\
40 & WD 0648+368   &  36.2 (1.0) &  31.8 (17.4) & 1 &   5.7 (5.6, 14.1) &  28.6 (28.2, 36.0) & old WD\\

\end{tabular}
\end{center}

\tablefoottext{a}{Number of photometric measurements used to compute the mean
  spectroscopic distance.}

\tablefoottext{b}{Numbers in parentheses are not rms errors but refer to the minimum
and maximum values where the distance of a given white dwarf is
varying within $d_{\rm spectro} \pm 1\sigma$}

\tablefoottext{c}{Summary of the status of the Hyades membership as discussed
  in this paper, including the {\it classical} members and the new candidates
  that are still valid {\it candidates} or that were rejected as {\it
    non-members} following our analysis. We also identify {\it old white
    dwarfs} for which the age is incompatible with a Hyades origin. Finally,
  one object has {\it uncertain} atmospheric parameters and we can not rule it
  out as a member. Note that binaries (three classical members, Nos. 1, 4 and
  7, and two new candidates, Nos. 12 and 13) are excluded from this table.}
  
\end{table*}

\section{Membership status of new Hyades candidates}

We have tested our $d_{\rm spectro}$ determinations with the classical Hyades,
and we expect these values to be equally precise for the new aspirants. In
this section, we review the status of the 17 new candidates (Nos. 11-27)
identified in Paper II and the 3 new candidates (Nos. 38-40) identified in
this work. For that purpose, we have computed in Table~2 the new distances
between the objects and the Hyades center ($d_{\rm Hyades~center}$) based on
the $d_{\rm spectro}$ values. It is clear that all but one (No. 20) of the new
candidates are outside of the tidal radius.

\begin{figure}[h!]
\begin{center}
\includegraphics[bb=10 250 602 632,width=3.6in]{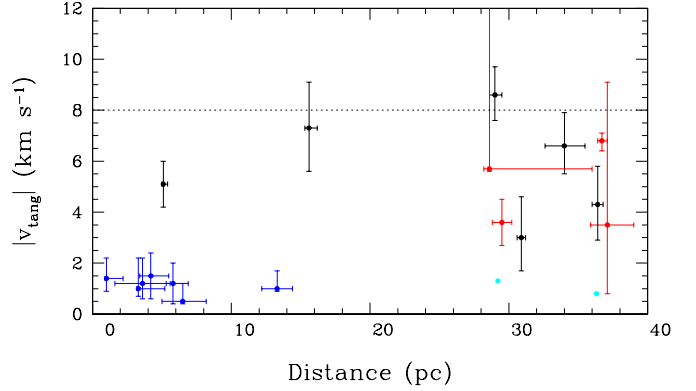}
\caption{Residual tangential velocity $\vert v_{\rm tang} \vert$ with respect
  to the Hyades group motion as a function of the distance to the cluster
  center, for classical members (blue points and error bars) and new
  candidates with an age lower (black points and error bars) and higher (red
  points and error bars) than the Hyades age. Also shown are the two
  candidates in binaries (Nos. 12 and 13, cyan points in the figure) for which
  we have no spectroscopic parameters. Instead we rely on the kinematic
  values, providing lower limits on the residual velocities.  The error bars
  are not rms errors but refer to the minimum and maximum values where the
  distance of a given white dwarf is varying within $d_{\rm spectro} \pm
  1\sigma$. The horizontal dotted line represents the 8 km~s$^{-1}$ residual
  velocity upper limit for the candidate selection.
\label{fg:f6}}
\end{center}
\end{figure}

Another quantity in Table~2 derived from the $d_{\rm spectro}$ calculation is
the residual tangential velocity $\vert v_{\rm tang} \vert$ with respect to
the expected motion of the star towards the cluster convergent point. In
Fig.~\ref{fg:f6}, the residual velocity of the candidates is plotted versus
the distance to the cluster center. Our updated $\vert v_{\rm tang} \vert$
values are by definition higher than those found in Paper II, where the
difference in the proper motion of candidates and that of the Hyades where
minimized on a plane perpendicular to the line-of-sight, in order to derive
the kinematic distances. Also included in Fig.~\ref{fg:f6} are the two
unresolved binaries for which we have no atmospheric parameters (Nos. 12 and
13), and we use their kinematic parameters in order to have the complete
picture of possible members. Because the companion of these candidates follow
very well the Hyades main sequence in the absolute magnitude vs. color diagram
for near-infrared colors (see Fig. 4 of Paper I), their kinematic distances
are expected to be fairly accurate.

According to Fig.~\ref{fg:f4}, the $d_{\rm spectro}$ and $d_{\rm kin}$
determinations do not agree within $1\sigma$ for most candidates, which would
imply that their membership is doubtful. On the other hand, if we allow them
to have residual velocities of a few kilometers per second, they could still
be valid unbound constituents. As a consequence, the residual velocities of
Fig.~\ref{fg:f6} are better suited to look at the membership status of new
candidates. The residual velocities are of the order of $\sim$1~km~s$^{-1}$
for the classical members, and one could in principle use this characteristic
value for a membership confirmation.  However, Paper I has shown that unbound
main-sequence stars had a significantly higher residual velocity dispersion
than the bound members, and we can expect a similar behaviour for white
dwarfs. Then, a higher residual velocity can be allowed to judge the status of
the new candidates.

It is already clear that 6 of the candidates (star Nos. 14, 16, 17, 18, 22 and
25) must be ruled out as Hyades components, since their updated distances
imply residual velocities of more than 20~km~s$^{-1}$. Indeed, it is obvious
that a cluster candidate can not have a residual velocity of the order of the
velocity dispersion of field white dwarfs, which is $\sim$30~km~s$^{-1}$ in
each spatial direction \citep{sion09}.

\begin{figure}[h!]
\begin{center}
\includegraphics[bb=10 250 622 632,width=3.6in]{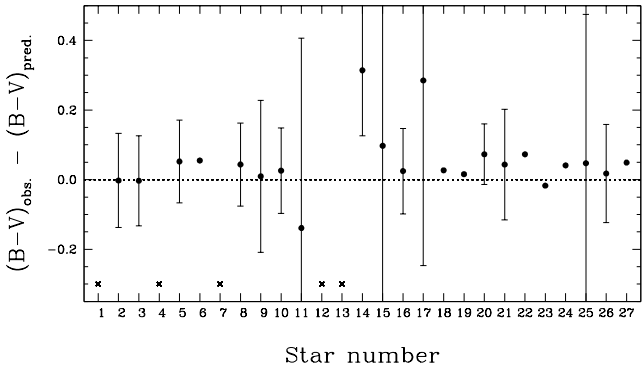}
\caption{Difference in the observed and predicted $B$-$V$ colors for the
  Hyades white dwarf candidates identified by their star number. The dotted
  horizontal line represents a perfect match between both quantities. Objects
  with a cross above the star No. are binaries.
\label{fg:f7}}
\end{center}
\end{figure}

Before we look at the remaining candidates, it is desirable to understand why
the above 6 background objects contaminated the sample of Paper II. From the
kinematic selection alone, it is clear that background, high velocity objects,
could pollute the sample. However, these background stars should have been
removed in the photometric selection, where more distant objects appear
subluminous compared to true Hyades members (see Fig.~\ref{fg:f1}).  In
Fig.~\ref{fg:f7}, we compare the observed and predicted $B-V$ colors, the
latter computed from model spectra with the atmospheric parameters from Table
1. We conclude that only two objects (Nos. 14 and 17) have been misidentified
as candidates since their published $B-V$ colors erroneously put them on the
white dwarf sequences. For the other four candidates rejected as members, we
conclude that they just happened to be {\it within} the theoretical sequences
by chance, but with a vertical offset. Indeed, for hot white dwarfs, the
absolute $V$ magnitude is rather insensitive to the $B-V$ color as can been
seen from Fig.~\ref{fg:f1}. More precisely, for a given color, a range of
$M_V$ of as much as two magnitudes is possible. It implies that it is very
difficult, from an absolute magnitude vs. color diagram alone, to conclude if
a kinematic candidate is a Hyades member, or a hotter and more distant
background white dwarf. The method of comparing kinematic and spectroscopic
predictions used in this work is then necessary to obtain a clean sample. One
might be intrigued by the fact that we find 6 background objects (up to 172 pc
from the Sun), and no foreground objects. The main reason seems that the
volume surveyed in this work in front of the cluster is only $\sim$1/50 of the
volume beyond the cluster up to the most distant background object. This
volume effect may not be important since catalogues of white dwarfs are far
from complete at distances greater than the Hyades center. On the other hand,
all of our background objects are hot DAs, and these objects are fairly rare
up to a radius of 20 pc \citep{giammichele12} and are more probably found at
larger distances.

There remain 14 objects, including 2 binaries with no spectroscopic
parameters, that are still possible members. Five candidates (Nos. 11, 15, 21,
39 and 40), however, are significantly older than the age of the cluster,
including one object (No. 15) that is not spectroscopically observed and hence
confirmed as a white dwarf.  Furthermore, we have one magnetic object with
imprecise atmospheric parameters (No. 27).  According to Fig.~\ref{fg:f6},
most of the remaining candidates have significant residual velocities of the
order of 5 km~s$^{-1}$ and lie far away outside of the tidal radius. These
candidates, if they are of Hyades origin, are not clearly linked to the
classical members in terms of their cluster center distances and residual
velocities, but could be escaped former members.

In principle, the diagram of Fig.~\ref{fg:f6} should be fairly complete up to
8 km~s$^{-1}$, although it must be kept in mind that not all white dwarfs
around the Hyades have been identified in white dwarf catalogues. The goals of
the next sections will be to understand the contamination from field white
dwarfs and the completeness of the white dwarf catalogues.

\subsection{Field white dwarfs contamination}

We made Monte Carlo simulations to estimate the number of field white dwarfs
that could lie within a radius of 40 pc of the cluster and mimic, by chance,
the Hyades motion, hence creating false positive detections. As input, we need
the luminosity function of white dwarfs within $\sim$100 pc of the Sun. Such a
function is not known, but, based on improved atmospheric parameters of
degenerate stars, \citet{giammichele12} published the luminosity function of
the relatively complete sample of white dwarfs within a 20 pc radius from the
Sun. In principle, it would be appropriate to extrapolate their results to the
Hyades region. However, the hot end of the luminosity function, which is of
significant interest for the young cluster, suffers from low number statistics
in the local sample of white dwarfs. Consequently, we rely on the SDSS
luminosity function \citep{harris06}, which includes a much larger number of
hot objects, but is incomplete by as much as 50$\%$ and complex completeness
corrections must be accounted for. Nevertheless, Fig.~22 of
\citet{giammichele12} shows that the local sample and SDSS luminosity
functions agree relatively well, and in particular, the derived total space
densities are very similar. Therefore, we simulated white dwarf populations
weighted by the SDSS luminosity function in a sphere of 40 pc radius centered
on the Hyades. The objects are randomly distributed in the galactic X, Y, Z
coordinates, with axes pointing to the Galactic Center (X), the direction of
galactic rotation (Y), and the North Galactic Pole (Z). The U, V, W velocities
are distributed according to a Gaussian weight function using the observed
velocity distribution from \citet{sion09} for the local 20 pc sample. They
find mean velocities, with respect to the Sun, of [$<$$U$$>$, $<$$V$$>$,
  $<$$W$$>$] = [$-$0.9, $-$21.5, $-$5.2] km~s$^{-1}$ and dispersions of
[$\sigma_U$, $\sigma_V$, $\sigma_W$] = [35.3, 33.1, 28.1] km~s$^{-1}$. To
derive the mean velocity and the velocity dispersion, \citet{sion09} assumed
zero radial velocity, which, at least, sets a lower limit to the velocity
dispersion.

\begin{figure}[h!]
\begin{center}
\includegraphics[bb=10 130 602 732,width=3.6in]{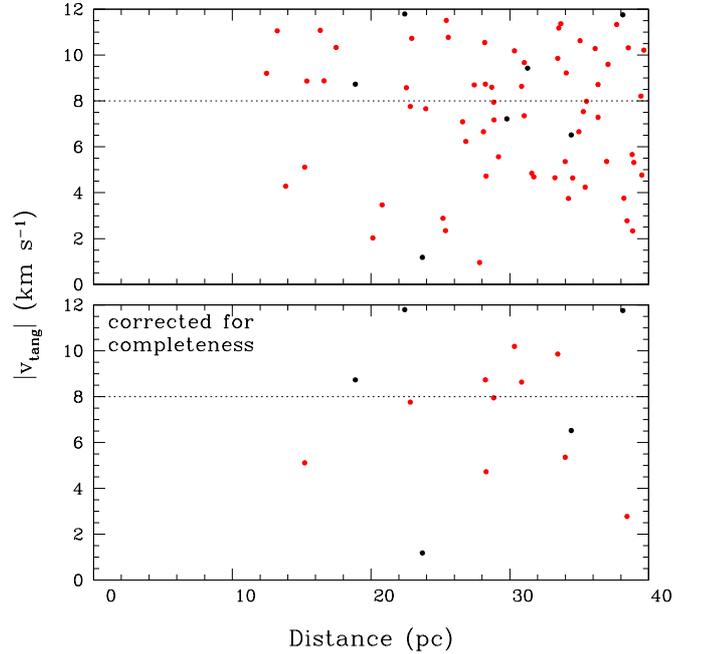}
\caption{{\it Top:} Residual tangential velocity with respect to the Hyades
  group motion, as a function of the distance to the cluster center, for a
  selected Monte Carlo simulation of field white dwarfs as described in the
  text. The red points are cool white dwarfs with $M_V > 12$. The horizontal
  dotted line represents the residual velocity upper limit of 8 km~s$^{-1}$
  used in Section 5.1. {\it Bottom:} Same simulation as the top panel, but
  corrected for the completeness of the WDC (see Section 5.3).
\label{fg:f9}}
\end{center}
\end{figure}

Our 100 simulations yield on average 1233 white dwarfs within 40 pc of the
cluster center. Nonetheless, we do not expect that many of these objects will
have velocities close to the Hyades group motion of [$-$41.1, $-$19.2, $-$1.4]
km~s$^{-1}$ (Paper I). On average, we find that 2.6 objects have 3D velocities
that are consistent with the Hyades motion within 8 km~s$^{-1}$. However, 3D
velocities can not be measured in most cases due to the lack of radial
velocities. Hence we also measured the residual 2D velocities of our simulated
objects with respect to the Hyades motion perpendicular to the
line-of-sight. In that case we find on average 29 objects that mimic the
cluster tangential motion within 8 km~s$^{-1}$. We show one of our simulations
in the upper panel of Fig.~\ref{fg:f9}, with the relative 2D velocity as a
function of the cluster center distance. The lack of objects close to the
center is simply a volume effect. We note that cool white dwarfs dominate the
luminosity function and also the local white dwarf sample, and consequently
88$\%$ of our simulated candidates are cool objects ($M_V >$ 12, red points on
the figure), for which the age, above $\sim$500 Myr, would be inconsistent
with a Hyades membership. In other words, only about 3 of the contaminants
would have an age consistent with the Hyades age. The summary of the predicted
number of polluting objects is given in Table~3. Finally, it should be
realized that due to the uncertainties in the observed luminosity functions
and velocity distributions, the numbers given in this section are only
accurate to about the 10$\%$ level.

The selection criteria of candidates in Paper II include a cross-match with
the White Dwarf Calatog (WDC) as well as with a few other sources. These
catalogues suffer from incompleteness and we have to take this into account in
our Monte Carlo simulations. In the following, we therefore investigate the
incompleteness of the WDC to determine a realistic estimate of the field star
contamination in our sample.

 \begin{table}[!h]
 \begin{center}
 \caption{Contamination from field white dwarfs}
 \label{tab3}
 \begin{tabular}{llll}
\hline \hline Residual velocity ($<$ 8 km s$^{-1}$) & Total & $M_V < 12$ & $M_V > 12$ \\ 

\hline

uncorrected for completeness & &\\
\hline
 3D & 2.6 & 0.30 & 2.3\\
 2D & 29  & 3.3  & 26 \\
\hline
corrected for completeness & &\\
\hline
 3D & 0.34 & 0.21 & 0.13 \\
 2D & 3.9  & 2.4 &  1.5 \\

\end{tabular}

\end{center}

\tablefoot{Given a maximum cluster center distance of 40 pc.}
  
\end{table}

\subsection{Completeness corrections}

\begin{figure}[h!]
\begin{center}
\includegraphics[bb=20 130 542 732,width=3.6in]{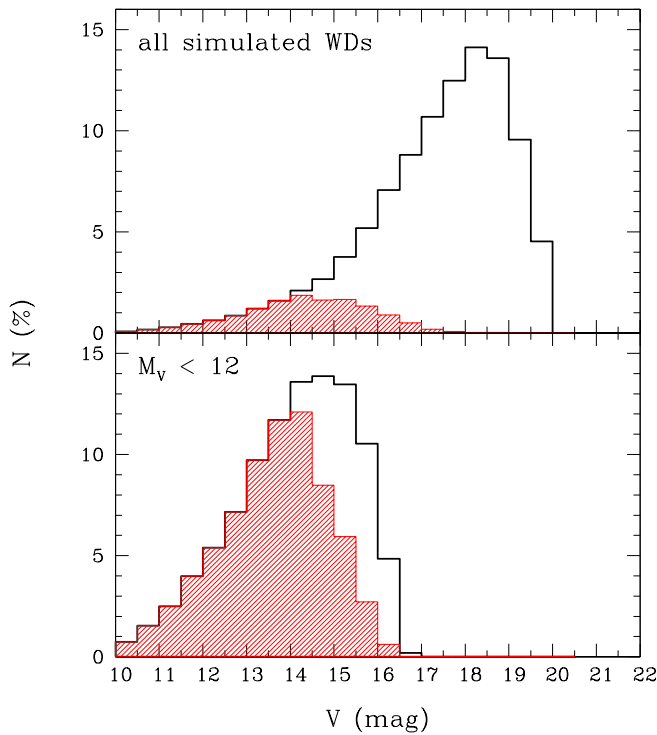}
\caption{{\it Top:} Distribution of $V$ magnitudes for the simulations of
  field white dwarfs discussed in Section 5.2 (bold histogram). The
  distribution has then been folded with our estimated completeness fraction
  of the WDC (red hatched histogram). {\it Bottom:} Same as top
  panel but only for objects with $M_V < 12$.
\label{fg:f10}}
\end{center}
\end{figure}

The $V$ magnitude distribution of the simulated white dwarfs from the previous
section is shown in Fig.~\ref{fg:f10} (top panel, bold histogram). Since the
local sample of white dwarfs is only complete up to about 20 pc and that the
objects in our simulations are on average twice as distant, it is already
obvious that many of them may be missing in white dwarf catalogues. One
important thing to notice, however, is that all of the faint objects with $V >
17$ in our simulations are cool white dwarfs with $M_V > 12$. Indeed, for
white dwarfs with $M_V < 12$, i.e. those with an age roughly consistent with
the age of the Hyades, the bottom panel of Fig.~\ref{fg:f10} shows that they
are always relatively bright ($V < 17$).

\begin{figure}[h!]
\begin{center}\includegraphics[bb=60 240 592 632,width=4.0in]{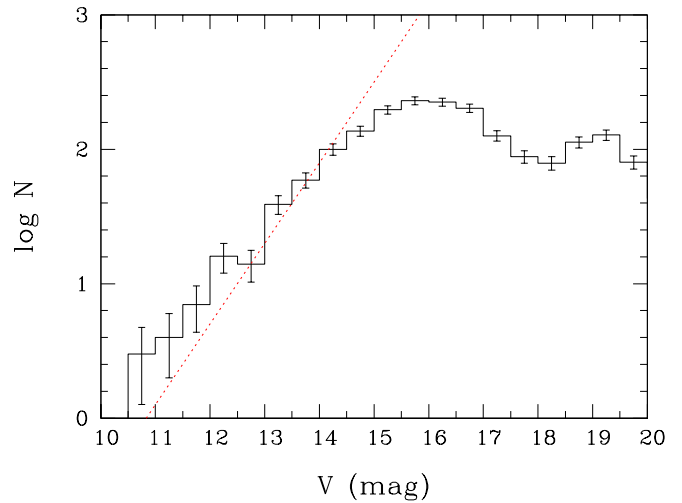}
\caption{Logarithm of the number of stars in the WDC as a
  function of their observed visual magnitude, in bins of 0.5 mag. The error bars
  are from number statistics. A linear
  relation assuming a volume complete sample (dotted red line) as been fitted to
  the data in the range $12.5 < V < 14.5$.
\label{fg:f11}}
\end{center}
\end{figure}

To estimate the completeness of the White Dwarf Catalog, we used the following
procedure. We computed the $V$ magnitude histogram for all objects in the
catalog\footnote{We have neglected magnitudes taken prior to 1972,
  photographic magnitudes, and those flagged as uncertain.}, which is
presented in Fig.~\ref{fg:f11} in terms of the logarithm of the number. We
neglected all objects first discovered in the SDSS. Removing these objects
seems justified to estimate the contamination of the Hyades sample, since all
candidates in Table~2 were known before the SDSS era, and SDSS only partly
covers the Hyades area. For uniform stellar density, a linear relation, with a
slope of 0.6, is expected if a sample is volume complete. We can see that the
observed distribution in the range $12.5 < V < 14.5$ is fitted well by the
expected relation. We computed the completeness at $V > 14.5 $ from the ratio
of observed to predicted objects. Afterwards, we folded this completeness
function with the distribution of $V$ magnitudes in our field white dwarfs
simulations presented in Fig. 10 (red dashed histograms). We derive from the
new total number of objects that only 14$\%$ of the field white dwarfs within
a radius of 40 pc to the Hyades center are found in the WDC. This fraction
increases to 72$\%$ if we only account for young white dwarfs with $M_V < 12$,
which implies that the completeness for old stars above this threshold is only
about 6\%. To find the number of field white dwarfs mimicking Hyades members,
we simply have to multiply the completeness fractions to the numbers of the
previous section. The result is given in Table~3, and, as an example, in Fig.
\ref{fg:f9} (bottom panel) where we modified the simulation by allowing
objects to randomly disappear in accordance with the completeness correction
as a function of magnitude. Hence, we predict that only 0.3 field white
dwarfs, on average, will be found within 8 km s$^{-1}$ from the 3D cluster
velocity.  Furthermore, 3.9 objects, among those 2.4 with $M_V < 12$ will
mimic the 2D velocity within the same range.

Although the relative larger number of white dwarfs at $V < 12.5$ in
Fig.~\ref{fg:f11} can be explained by statistical fluctuations, they can also
indicate that our assumption of a completeness of 100\% up to $V = 14.5$ is
rather optimistic\footnote{Note that for $V<11$, most objects are white dwarfs
  in Sirius-like systems, for which the completeness is rather low
  \citep{holberg09}.}. A larger completeness correction provides a lower
number of field white dwarfs contaminating our sample of Hyades
candidates. Therefore the contamination estimates above can be considered as
conservative.

\section{Discussion}

Our analysis has so far reviewed the status of the ten classical Hyades white
dwarfs and provided improved distances based on the most recent determinations
of the atmospheric parameters. Then, we updated the residual velocities with
respect to the bulk motion of the cluster. We have furthermore studied 20 new
candidates identified in Paper II and in this work. We found that all but one
were outside of the tidal radius, and six of them were clearly background
objects. There remain 14 objects for which the Hyades membership status is
still possible. A comparison of the distribution of the observed candidates
versus the simulated field white dwarfs in Fig.~\ref{fg:f6} and
Fig.~\ref{fg:f9} (bottom panel, with the completeness correction) can serve to
conclude about the global status of those remaining candidates.

We will first draw our attention on objects that have an age lower than the
cluster age (black points in Fig.~\ref{fg:f6}), which are the most logical
candidates. There are six such white dwarfs, although this excludes the two
binary candidates for which the cooling age is not known (cyan points). Also,
the magnetic white dwarf WD 0816+376 is at 47 pc from the cluster center
according to Table~2, but given the imprecise atmospheric parameters, we can
not exclude that it is actually closer (see Table~2). In comparison, the
number of contaminants in this age group is 2.4 or 3.3, including or not the
completeness correction, respectively. We find that the probability that six
objects are field stars is 4\% or 12\% for a Poisson distribution including or
not the completeness correction, respectively.  Compared to the field, there
is an overdensity of white dwarfs of ages younger than the Hyades. We must
remind, however, the difficulty of estimating the uncertainty on the number of
contaminants, although we believe that our proposed number is conservative.
  
If we look closer at the properties of the 6 most promising candidates
(identified in Table~2 by their {\it candidate} status), we can see from Table
1 that their masses are in the range of 0.66 $< M_{\odot} <$ 0.98, with a mean
of 0.82 $M_{\odot}$, i.e. much higher than the mean mass of field white
dwarfs. The probability of 6 field stars having such a high mean mass,
assuming the volume-complete mass distribution of \citet{giammichele12}, is
only $\sim$0.1\%. Once again, these 6 objects stand out from the field white
dwarfs.

Secondly, we can look at older stars ($M_V >$ 12, red points in the
Fig.~\ref{fg:f6} and \ref{fg:f9}) for which we have four candidates. For the
time being, No. 15 remains an additional candidate, given its uncertain nature
and imprecise atmospheric parameters. From our simulations, we expect between
1.5 and 26 old objects to contaminate the sample, considering or not the
completeness correction, respectively. Since the completeness correction is so
high, it is hard to be confident about the expected number of contaminants.
Taking literally the mean of 1.5 contaminants, the probability to find 4 is
only 7\%. However, one would expect that these are the most obvious
contaminants, since their ages are inconsistent with the Hyades age.

Nevertheless, an important problem arises if one concludes too fast that the
old white dwarfs are contaminants. Indeed, two of these candidates have radial
velocities that are in agreement with the Hyades motion (WD~0120$-$024;
\citealt{silvestri01}, and WD~0433+270; \citealt{zuckerman03}), which would
make them strong candidates if it was not for their age of about 4 Gyr (see
Table~1). These two objects are in common proper motion pairs with
main-sequence stars considered as of Hyades origin. We note, however, that
\cite{silvestri01} also measured a radial velocity for WD~0433+270, which is
incompatible with the value found by \citet{zuckerman03}, and also
incompatible with the radial velocity of its companion, since the mass, hence
the gravitational redshift correction, is known for this white dwarf. Our
simulations show that we expect to find, on average, 2.3 old objects that
mimic the 3D Hyades velocities within 8 km~s$^{-1}$, although when accounting
for the completeness correction, the number drops to about 0.1. The fact that
we find two of them may be statistically improbable, but not impossible
considering low number statistics.  If we instead infer that they are not
simply field stars that accidentally have the 3D Hyades motion, different
scenarios can be conceived. Perhaps the most promising scenario is that these
objects could be members of the Hyades supercluster cold kinematic stream
which is known to have components that are significantly older than the Hyades
\citep{chereul99,famaey07}. Other possibilities are that they could have been
captured by the Hyades cluster, or they could be younger Fe-core white dwarfs
\citep{catalan08}, although in the latter case, the authors claim that the
white dwarfs would still be too old ($\sim$1 Gyr).

As a consequence of the discussion above, further observations are needed to
constrain the membership of the Hyades candidates. Accurate trigonometric
parallaxes, for instance from Gaia, would provide distances that are
independent of $d_{\rm spectro}$ and therefore can eliminate possible
uncertainties due to the atmospheric parameters. This alone, however, is not
sufficient to conclude about their membership, since their true 3D velocities
are not known. Consequently, accurate radial velocity measurements are
indispensable to confirm or rule out which of the candidates is of Hyades
origin. If consistent with the predicted Hyades radial velocity, the
probability of contamination is very low (see Table~3). Due to the extreme
broadening of the spectral lines, the spectroscopic measurement of the radial
velocity of white dwarfs is extremely difficult to obtain, and only possible
with the help of the narrow NLTE cores of H$_\alpha$
\citep{greenstein77,koester87,falcon10}. Since this measurement must be
corrected for the gravitational redshift, typically $\sim$70 km~s$^{-1}$, such
a determination is only possible if we have a good knowledge of the mass. For
instance, star No. 2 has a radial velocity measurement that is not in
agreement with the Hyades motion (see Paper II), even though it is still
considered as a classical member. It is more likely, in this case, that the
uncertainty of the radial velocity measurement is underestimated.  One should
therefore be cautious about false negative conclusions from radial velocities
alone.

In principle, there is a completely different way to measure the radial
velocity astrometrically by determining the perspective acceleration due to
the changing distance of the white dwarfs \citep{oort32,gatewood74} with Gaia
\citep{anglada10}. In this approach no correction for gravitational redshift
is needed. Assuming that a 3$\sigma$ deviation of the proper motion during the
5 years mission is required and taking into account that the Gaia
proper-motion standard error depends on the magnitude \citep{debruijne12}, one
concludes that only bright ($ V < $ 13) white dwarfs closer than 5 pc allow an
estimate of the radial velocity with the needed precision (de Bruijne,
priv. comm.).  Hence we can not rely on this method for the Hyades.

\subsection{Implications in terms of the Hyades evolution}

This paper has performed a spectroscopic check of the 27 white dwarfs found in
Paper II as possible Hyades members. Compared to Paper II, we tried to reveal
more kinematic candidates by softening the velocity criterion. Nevertheless,
no more than 10 white dwarfs have been found within the present-day tidal
radius (9 pc) of the cluster, including one new candidate (No. 20). Also, no
objects with an age between 340 and 600 Myr are recovered in this radius
range. Here we claim that no more single white dwarfs of ages lower than the
Hyades age will be detected within the tidal radius. Outside the tidal radius
and up to 40 pc, between 6 and 9 white dwarfs with a possible Hyades origin
are found. Among those, about three are expected to be contaminants, although
a few hot degenerate candidates could also be missing due to the
incompleteness of the white dwarf catalogues. Even among the 6 most promising
new candidates, all have an age lower than 340 Myr, except for No. 26, with an
age of 410 Myr. In summary, we face again the long-standing problem
\citep{tinsley74,weidemann77} that not enough old white dwarfs are associated
with the cluster in comparison to predictions, in particular compared to the
most recent N-body Hyades cluster simulations of \citet{ernst11}.

The simulations of \citet{ernst11} use the present-day mass distribution of
stars within 9 pc of the cluster center from Paper I among inputs (see their
Table~1) to simulate the formation and evolution of the cluster. Their best
fit simulations ultimately predict the number, mass, and age of white dwarf
members. They forecast that 73 remnants were formed in the cluster so far,
including five black holes, 13 neutron stars and 55 white dwarfs. However,
many of these remnants have long ago evaporated from the cluster. In the
simulations, the newly formed white dwarfs get a kick drawn from a Maxwellian
distribution with a 1D velocity dispersion of 5 km s$^{-1}$, corresponding to
a 3D kick with a mean of $\sim$8 km s$^{-1}$. The present-day consequence of
these kicks is that 40\% (22 objects) of the white dwarfs remain within the
present-day tidal radius and are dynamically cooled down to show a velocity
dispersion of only $\sim$0.5 km s$^{-1}$.  It is worth noting that only
$\sim$10 among these bound white dwarfs have ages smaller than 340 Myr (Ernst
et al., private communication). On the other hand, the kicks have removed 33
white dwarfs from the cluster. At present, five lie in the range $ 9 < d_{\rm
  center} < 30$ pc, and others are more than 30 pc away from the center. The
authors are cautious to warn that the final number of tidally bound white
dwarfs depends critically on the kick velocity.  However, there are actually
few observational constraints to confirm if white dwarfs obtain such kicks
from asymmetrical mass loss in the planetary nebula ejection, and how strong
it would be \citep{davis08,fregeau09}.  We also note that possible encounters
with giant molecular clouds or spiral arms are neglected in the simulations.

Compared to the N-body simulations of \citet{ernst11}, the observations cannot
account for at least half of the predicted white dwarfs within the tidal
radius, but observations and simulations agree for the number of young ($
<$340 Myr) white dwarfs there. It is also worth to note that the simulations
predict only a small number (five) of white dwarfs between 9 and 30 pc, which
would be consistent with our findings of candidates in the 40 pc range. On the
other hand, the observations and simulations fully disagree for the older
white dwarfs: the simulations suggest at least ten older ($>$340 Myr) white
dwarfs within the tidal radius, but we do not observe any of them. It is very
unlikely that the ages of white dwarfs drawn from the cooling sequences of
\citet{wood95,fontaine01} are significantly underestimated given the
consistency between white dwarf cooling ages and main sequence ages in the
Hyades and other clusters \citep{degennaro09,jeffery11}. We have verified that
even with the calculations of \citet{renedo10}, which are done by evolving
white dwarf progenitors directly from the zero-age main sequence, the ages of
the white dwarfs discussed in this work are only changed by $\sim$10\%. There
are different other explanations for this discrepancy and we review some of
them in turn.

One possibility is that more white dwarfs than expected were kicked out of the
cluster, and that most of them are now far from the center. The kick
velocities added to white dwarfs in the \citet{ernst11} simulations can be
seen as some kind of free parameter. However, since objects within the tidal
radius thermally relax on short time scales, they can not be used to constrain
this parameter. The observed spatial density and residual velocities of white
dwarfs outside of the tidal radius could in principle be used, but these
numbers are poorly constrained by the current work. We suggest that new
simulations could be run with a wide number of assumptions on the kick
velocities, in order to predict how the number density and residual velocities
would be modified. \citet{kicks} suggest that older white dwarfs are
preferentially lost due to their higher recoil speed, which is in turn due to
the larger amount of mass loss of their high-mass progenitors, although it
only remains an hypothesis. We note that similar simulations for the white
dwarfs in other open clusters could be very helpful to constrain the
distribution of kick velocities.

Another scenario is that fewer white dwarfs than predicted were ever formed in
the Hyades, which would correspond to a change of the Initial Mass Function
(IMF) at its high-mass end.  \citet{dobbie06} estimate from their initial
vs. final mass relation that the classical white dwarfs come from
main-sequence progenitors in the range $\sim$2.5-4 $M_{\odot}$. Therefore, to
obtain more promising predictions from the simulations, one would have to
modify the formation rate of white dwarfs for main-sequence stars with $M > 4$
$M_{\odot}$.  Related to this, the fact that initial vs. final mass relation
for white dwarfs is not so well constrained for high masses \citep[][see their
  Fig. 3]{dobbie06} might also have an effect on the number of white dwarfs
vs. neutron stars and the age distribution of the remnants.  Having a steeper
initial mass function in this regime could have a strong effect on the
predicted number of white dwarfs \citep{bc07}. In this context we note that
the present-day mass function of the Hyades derived in Paper I (Fig. 10)
tolerates a steeper slope leading to less white dwarf progenitors.

Thirdly, the topic of white dwarfs in binary systems has only been touched
marginally in this paper. \citet{holberg09} claims that 32 $\pm$ 8 \% of the
local white dwarfs are in binary systems. Among the classical members, we find
that three out of ten are binaries, a number similar to field
objects. However, it would be dangerous to extrapolate that the fraction of
binaries is the same for field white dwarfs and the Hyades. \citet{sollima10}
estimate that the binary fraction of stars in open clusters is between 35 and
70\%.  Based on the Hipparcos data from \citet{age}, \citet{bc07} found that
the binary fraction in the Hyades increases from 30\% for late-G to 88\% for
early-A stars.  While we claimed previously that virtually all single white
dwarfs have been discovered in the cluster center area, it is likely not the
case for white dwarfs in unresolved binaries. For instance, V471 Tau and HD
27483 (Nos. 1 and 4) have not been discovered by systematic surveys, but from
specific studies of Hyades main-sequence members \citep{nelson70,bc93}. Since
old and massive Hyades white dwarfs (e.g. with masses of $\sim$0.8
$M_{\odot}$) are completely hidden in the visible by their main-sequence
companion in the majority of cases, they can only be detected by UV
observations of each main-sequence member.  However, \citet{bc95} found no
white dwarf companion in a sample of IUE spectroscopic observations for 27
Hyades F stars. Furthermore, \citet{paulson04} found no evidence of white
dwarf companions in a radial velocity survey of 94 Hyades stars. All in all,
there is so far no observational evidence that the missing white dwarfs older
than $\sim$340 Myr are hidden in binary systems.

\section{Summary}

The starting point of our work has been the Hyades white dwarf candidates
kinematically and photometrically selected by \citet{schilbach12} from the
PPMXL survey. We evaluated the membership status of these candidates by using
spectroscopic and photometric observations, from which we determined new
distances of the white dwarfs from the Sun and from the cluster center as well
as residual velocities with respect to the Hyades bulk motion. We find a very
good agreement between kinematic and spectroscopic distance predictions for
the 10 classical white dwarf members, confirming that the kinematic and
photometric data sets, as well as model atmospheres that we rely on, are
consistent with each others.

We ruled out 6 of the Hyades candidates from Paper II as background stars, but
also added 3 new objects by increasing the residual velocity upper limit for
membership selection to 8 km s$^{-1}$. There remain 14 candidates, five of
which have cooling ages higher than the age of the cluster. We do not consider
them as of Hyades origin. For another three, the atmospheric parameters are
rather imprecise or could not be determined, leaving us with six
spectroscopically confirmed hot and young white dwarfs consistent with a
Hyades membership. In order to estimate the contamination fraction from field
white dwarfs, we performed Monte Carlo simulations.  The result is that
$\sim$2-3 hot and young white dwarfs may contaminate our sample yielding that
the probability that all six are contaminants is low. A final confirmation of
the candidates needs accurate radial velocity measurements properly corrected
for gravitational redshift.


The major discrepancy between observations of white dwarfs of Hyades origin
and the predictions by simulations is the lack of about a dozen white dwarfs
with cooling ages between 340 and $\sim$600 Myr, which should be found within
the present-day tidal radius ($\sim$9 pc) of the cluster.  Possible
explanations for this are: a) at their formation, within the first 300~Myr of
the cluster's life, white dwarfs get a large ($\gtrsim$5 km s$^{-1}$) kick
velocity relative to the cluster motion which sweeps them farther out than
40pc from the centre, b) the high-mass end of the IMF is steeper than the
Salpeter slope of 2.3 assumed in the simulations, hence leading to a lesser
number of massive stars, and c) there is still a number of so-far unrevealed
white dwarfs hidden in binary systems. There seems to be no way, at present,
to decide which of these scenarios is the dominant one in consistence with the
conclusion of \citet{williams04}.

\begin{acknowledgements}

We wish to dedicate this work to the memory of Volker Weidemann who was very
fond of the Hyades white dwarfs. P.-E. T. is supported by the Alexander von
Humboldt Foundation. This work was partially supported by
Sonderforschungsbereich SFB 881 "The Milky Way System" (Subproject A4 and B5)
of the German Research Foundation (DFG).

\end{acknowledgements}

\bibliographystyle{aa} 

\end{document}